\documentclass[twocolumn]{aastex62}

	\usepackage{graphicx}
	\usepackage{float}
	\usepackage{savesym}
	\savesymbol{tablenum}

	\usepackage{amsmath}

	\usepackage{gensymb}

	\usepackage{siunitx}
	\DeclareSIUnit\groove{(grooves)}

\begin{document}

\title{Thermal instability of thin accretion disks in the presence of wind and toroidal magnetic field}
\shorttitle{Thermal instability of thin accretion disks ...}
\author{Asiyeh Habibi}
\author{Shahram Abbassi}
\affil{Department of Physics, School of Sciences, Ferdowsi University of Mashhad, Mashhad, 91775-1436, Iran;  \textcolor{blue}{abbassi@um.ac.ir}}
\begin{abstract}
We study the local thermal stability of thin accretion disks. We present a full stability analysis in the presence of a magnetic field and more importantly wind. For wind, we use a general model suitable for adequately describing several kinds of winds. First, we explicitly show that the magnetic field, irrespective of the type of wind, has a stabilizing effect. This is also true when there is no wind. In this case, we confirm the other works already presented in the literature. However, our main objective is to investigate the local thermal stability of the disk in the presence of the wind. In this case, interestingly, the response of disk is directly related to the type of wind. In other words, in some cases, the wind can stabilize the disk. On the other hand, in some cases, it can destabilize the disk. We found that in some thin disk models where the magnetic pressure cannot explain the stability of the disk by including a typical contribution for magnetic pressure, the wind can provide a viable explanation for the thermal stability.
\end{abstract}
\section{Introduction}
In a pioneer work, \cite{1973A&A....24..337S} discovered the thin accretion disk model. This model is extremely successful to explain the physical properties of black hole X-ray binaries. However, after more than four decades there are still unsolved questions and puzzles in the structure of thin disks. For a comprehensive review of this model, we refer the reader to \cite{2014ARA&A..52..529Y}.  In this paper, we focus on one of these puzzles dealing with the local thermal stability of the disk. For mass accretion rates higher than a few percents of Eddington rate, the radiation-dominated thin disks get thermally unstable. This is a well-established fact widely investigated in the literature, for example, see \cite{1976MNRAS.175..613S} and \cite{1978ApJ...221..652P}. More realistic calculations, i.e., numerical simulations, show that the final fate of unstable regions is a nonlinear oscillation between two stable phases \cite{2007ApJ...666..368L}. According to these theoretical studies, in real observations, one may expect a substantial variability in the physical properties of black hole X-ray binaries.

However, the high/soft state of X-ray binaries appears to be quite stable on observation \cite{2004MNRAS.347..885G}. These observations reveal a little variability with luminosities ranging from 0.01 to 0.5 $L_{Edd}$ which directly means that this thin-disk configuration is thermally stable. This conflicts with the accretion disk theory. Generally there are two processes which may likely change this inconsistency between theory and observations. The first scenario is that the viscose stress is proportional to gas pressure instead of total pressure, which in this case disk will be stable against thermal instability, \cite{1981ApJ...247...19S}. The second process is considering a mechanism for making disk cooler which may eliminate instability or equivalently increasing the relative importance of gas pressure in comparison of radiation pressure. To address this process several investigations have been proposed. \cite{1994ApJ...436..599S} have shown that instabilities will remove if the most gravitational energy released in the disk is transported to the corona. Convective cooling has been suggested as a stabilizing mechanized by \cite{1995ApJ...443..187G} which some later investigations have been pointed out that it might have a minor effect on the disk instability. \cite{2013MNRAS.434.2262Z} considered possible effect of turbulence on the instability of the disks. Cooling the disk with when the magnetic pressure becomes important in the hydrodynamical equilibrium (\cite{yuan}) or cooling via the dynamically or magnetically driven wind \citep{2014ApJ...786....6L} are the other possible solutions. However, it is necessary to mention that the puzzle has not been resolved yet. As one of the last suggested solutions, \cite{yuan}, claimed that the existence of magnetic pressure in the system may substantially stabilize the disk. However, one needs a substantial fraction for the magnetic pressure to achieve stability. More specifically, for the stability, the magnetic pressure should contribute more than $20$\% of the total pressure. This fraction is too large compared with the typical values common in the simulations. 

Our paper is close in spirit to \cite{yuan}. We revisit the stability problem by generalizing the linear analysis introduced in \cite{yuan} by adding a wind mechanism. The importance of wind/outflow in angular momentum removal from many accreting systems is supported by strong observational evidence,  e.g. \citep{2005Natur.435..652W}. On the other hand, it was long apparent that a disk wind/outflow contributes to loss of mass, angular momentum, and thermal energy from accretion disks in theoretical modeling, e.g. \citep{1982MNRAS.199..883B}. To add wind/outflow effect we use a parametric simple model presented by \cite{knigge} which derived the radial distribution of the dissipation rate and effective temperature across a Keplerian, steady-state, mass-losing accretion disk, using a simple parametric approach. This simple model is sufficiently general to be applicable to many types of wind like radiation driven outflow and centrifugally driven wind. Using this model, we show that mass loss via wind can stabilize the disk. 

The outline of this paper is as follows: in section \ref{bequations} we write the basic equations governing the system in the presence of magnetic field and mass loss via wind. In section \ref{ins} we investigate the local thermal stability of the disk and find a new criterion for the stability. This criterion is examined in different situations and the result is presented in \ref{Result}. Finally, conclusion and discussion are presented in section \ref{sum}.


\section{Hydrodynamics equations in the presence of wind and magnetic field}
\label{bequations}
In this section, we introduce the basic equations governing the dynamics of our accretion thin disk model. More specifically, to find a criterion for the thermal stability of the disk, we use the conservation equations of energy, mass and angular momentum in the presence of wind and a toroidal magnetic field. Conveniently, we choose the cylindrical coordinate system $(R,\phi,z)$. In our simple model, the disk consists of a differentially rotating disk around a central mass $M$. For the sake of simplicity, we assume that the flow is static and axisymmetric, i.e., $\frac{\partial}{\partial t}=0$ and $\frac{\partial}{\partial\phi}=0$.

Assuming that the disk lies in the $x-y$ plane, the hydrostatic condition in the vertical direction $z$ takes the following form
\begin{equation}
\frac{\partial p}{\partial z}+\rho \frac{\partial \psi}{\partial z}=0
\label{hydro}
\end{equation}
in which $p$ is the fluid pressure, $\rho$ is the mass density and $\psi$ is the Newtonian gravitational potential of the central mass. It should be mentioned that we restrict ourselves to regions far enough from the center of the disk, to ignore relativistic effects. Furthermore, we assume that the gravitational potential is dominated by the central mass and ignore the contribution of the disk itself. On the other hand, as we mentioned, we deal with a thin disk. In this case, one may simply assume that at every radius $r$, the vertical thickness is very small compared to $R$, i.e., $R/z \ll 1$. Keeping this approximation in mind, we can write
\begin{gather}
\frac{\partial p}{\partial z}\simeq - \frac{p_{\text{tot}}}{H} ,~~~~~
\frac{\partial \psi}{\partial z}\simeq  \frac{G M H}{R^3}
\label{hydro2}
\end{gather}
Where $p_{\text{tot}}$ is the total pressure in the midplane, $H$ is the disk scale height and $R$ is the radial coordinate. By total pressure we mean the combination of all pressure contributions, namely the gas pressure $p_{\text{gas}}$, radiation pressure $p_{\text{rad}}$ and the magnetic pressure $p_{\text{mag}}$. Therefore $p_{\text{tot}}$ reads 
\begin{equation}
p_{\text{tot}}= p_{\text{gas}}+p_{\text{rad}}+p_{\text{mag}}
\end{equation}
Hereafter we assume that the gas content of the disk is an ideal classic gas. On the other hand, the magnetic field is assumed to be toroidal, i.e., $\mathbf{B}\simeq B_{\phi} \hat{\mathbf{\phi}}$ where $\hat{\mathbf{\phi}}$ is the azimuthal unit vector. The pressure contributions can be written as 
\begin{equation}
p_{\text{gas}}=\frac{2 k_{B} \rho T}{m_{H}},
\label{ptot}
\end{equation}
\begin{equation}
p_{\text{rad}}= \frac{1}{3} a T^{4},
\end{equation}
\begin{equation}
p_{\text{mag}}= \frac{B_{\phi}^2}{8 \pi}
\end{equation}
where $k_{B}$ is Boltzmann constant, $m_{H}$ is hydrogen mass, $T$ is the temperature at midplane and $a$ is  the radiation constant. On the other hand in the thin disk limit, the surface density is written as $\Sigma=2\rho H$. Therefore it is straightforward to show that equation \eqref{hydro2} leads to the following relation for the total pressure
\begin{equation}
p_{\text{tot}}=\frac{M G H \Sigma(R)}{2 R^3}
\label{totp}
\end{equation}

Flow equations in the presence of wind can be easily obtained from Navier-Stocks equations, see \cite{knigge} for more details. The continuity equation reads
\begin{equation}
\frac{\partial}{\partial t}(2 \pi R \Sigma)- \frac{\partial \dot{M}_{\text{acc}}}{\partial R}+ \frac{\partial \dot{M_{\text{w}}}}{\partial R}=0
\label{cont}
\end{equation}
where $\dot{M_{\text{w}}}$ is the mass-loss rate from the disk caused by the wind. The mass-loss rate is related to the mass-loss rate per unit area, i.e., $\dot{m}_{\text{w}}(R)$, as follows
\begin{equation}
\dot{M_{\text{w}}}(R)=4\pi\int_{R_*}^{R}\dot{m}_{\text{w}}(R')R' dR'
\label{mw}
\end{equation}
where $R_*$ is the inner edge of the disk. In principle it can not be smaller than the radius of the central mass. In our analysis we consider it as a few times of the Schwarzschild radius $R_g=2 M_* G/c^2$. On the other hand, $\dot{M}$ is the mass accretion rate defined as $\dot{M}_{\text{acc}}=-2\pi R v_R \Sigma>0$, where $v_R<0$ is the radial inflow velocity. Since we have assumed that the flow is static, one may simply infer from equation \eqref{cont} that $-\dot{M}_{\text{acc}}(R)+\dot{M}_{\text{w}}(R)=const$. The mass-loss rate at $R=R_*$ is zero. Therefore the above mentioned constant is $\dot{M}_{\text{acc}}(R_*)$ and we have
\begin{equation}
-\dot{M}_{\text{acc}}(R)+\dot{M}_{\text{w}}(R)=\dot{M}_{\text{acc}}(R_*)
\label{mah1}
\end{equation}

Accordingly the angular momentum conservation for the thin disk can be obtained as follows \cite{knigge}
\begin{equation}
\begin{split}
\frac{\partial}{\partial t}(\Sigma R^2 \Omega)+&\frac{1}{R}\frac{\partial}{\partial R}(R^3 \Sigma v_r \Omega)=\\& \frac{1}{R}\frac{\partial}{\partial R}(R^3 \Sigma \nu \frac{\partial \Omega}{\partial R})-\frac{l^2 R^2\Omega}{2\pi R}\frac{\partial \dot{M_{\text{w}}}}{\partial R}
\end{split}
\label{av1}
\end{equation}
where $\nu$ is the kinematic viscosity. We have assumed that at radius $R$, the ejected matter through the wind, carries away an angular momentum per unit mass  $l^2 R^2 \Omega(R)$. This model leads to a simple approach suitable for many types of wind models, for more details on this model we refer the reader to \cite{knigge}. Conveniently, $\Omega(R)$ is the angular velocity of the flow at radius $R$ and $l$ is a dimensionless parameter useful to characterize the main properties of the wind. More specifically, $l=0$ indicates a non-rotating disk wind. On the other hand, $l=1$ corresponds to outflowing material that carries away the angular momentum it had at the radius of ejection \cite{knigge}. Similarly, $l>1$ belongs to centrifugally driven disk winds that remove a lot of angular momentum from the disk. 

We reiterate that we use Newtonian potential for the central mass and ignore the self-gravity of the disk. Consequently, the angular velocity is given by the Keplerian angular velocity $\Omega_K=\sqrt{M_* G/R^3}$. 

Let us first assume that there is no wind mechanism in the system. In this case, one may simply integrate equation \eqref{av1} over the radial range $(R_{\text{in}},R)$ to obtain
\begin{equation}
\dot{M}_{\text{acc}}(R)(\Omega_{K} R^2- l_{\text{in}})= 2\pi R^2 T_{R\phi}
\label{av2}
\end{equation}
in which $R_{\text{in}}=3 R_g$ indicates a region inside which there is no stable circular orbit, and $l_{\text{in}}=\sqrt{M_* G R_{\text{in}}}$ is the specific angular momentum of the last stable orbit.  On the other hand the $R\phi$ component of the energy-momentum tensor when the system is axisymmetric is given by $T_{R\phi}=\nu\Sigma R d \Omega/d R$. Now, let us modify equation \eqref{av2} in order to include the wind contribution to the distribution of angular momentum throughout the disk. To do so we multiply equation \eqref{av2} by $2\pi R$ and ignore the time derivatives. Using the continuity equation \eqref{cont}, and by integrating along the radial coordinate in the interval $(R_{\text{in}},R)$, we obtain
\begin{equation}
\dot{M}_{\text{acc}}(R)(\Omega_{K} R^2- l_{\text{in}})+C_\text{w}(R)= 2\pi R^2 T_{R\phi}
\label{av4}
\end{equation}
where $C_\text{w}(R)$ is the correction term induced by the existence of the wind in the system. This term is written as
\begin{equation}
C_\text{w}(R)=\frac{4\pi K(l^2-1)l_{\text{in}}}{\sqrt{R_*}(\xi+5/2)}(R^{\xi+5/2}-R_{\text{in}}^{\xi+5/2})
\label{av3}
\end{equation}
It should be emphasized that we have taken a simple power law model for the mass-loss rate per area as follows
\begin{equation}
\dot{m}_{\text{w}}(R)=K R^{\xi}
\end{equation}
where $K$ and $\xi$ are two free parameters \cite{knigge}. By setting $l=1$ we see that $C_\text{w}(R)$ vanishes and equation \eqref{av3} coincides with \eqref{av4}. Therefore it seems that the effects of wind disappear in the system. However one should note that although $C_\text{w}(R)=0$, the accretion rate $\dot{M}_{\text{acc}}$ in equation \eqref{av3} is a function of radius because of the wind. While when there is no wind, the accretion rate is constant everywhere throughout the disk.

The last governing equation is the energy equation. This equation in the presence of wind can be written as follows
\begin{equation}
Q^{+}_{\text{vis}}=Q^{-}_{\text{rad}}+Q^{-}_{\text{adv}}+Q^{-}_{\text{win}}
\label{q1}
\end{equation}
plus (minus) sign stands for physical processes that produce (suppress) heating in the system. The viscous heating reads
\begin{equation}
Q^{+}_{\text{vis}}=-T_{R\phi}R \frac{d\Omega}{d R}
\label{q2}
\end{equation}
before introducing the other energy components, let us briefly explain the specific form of $T_{R\phi}$ widely used to study the structure of thin and thick accretion disks. Both analytic and MHD simulations show that $T_{R\phi}$ is the dominant component of the stress tensor and one may model it as follows
\begin{equation}
T_{R\phi}=2\alpha p_{\text{tot}}H
\label{q3}
\end{equation}  
where $\alpha$ is the viscosity coefficient and plays a key role in the $\alpha$ viscosity model of accretion disks. For a short review on this choice, we refer the reader to \cite{yuan}. One may simply write $T_{R\phi}$ in terms of the surface pressure $P$ as $T_{R\phi}=\alpha P$. In this case, the dimension of $P$ is Newton per meter.

On the other hand the cooling via radiation is given by
\begin{equation}
Q^{-}_{\text{rad}}=\frac{32 \sigma T^4}{3\tau}
\end{equation}
where $\tau=\kappa \Sigma/2$ is the optical depth and $\kappa\simeq 0.4\, \text{cm}^2 \text{g}^{-1}$ if the opacity in the inner parts of the disk is mostly due to electron scattering. Furthermore, $\sigma$ is the Stefan-Boltzmann constant and $T$ is the temperature of the disk.

The advection cooling in the disk is related to accretion rate as follows \cite{abbromo}
\begin{equation}
Q^{-}_{\text{adv}}=\mu\frac{\dot{M}_{\text{acc}} (R)\Omega_K^2 H^2}{2\pi R^2}
\end{equation}
we use the typical value $\mu=1.5$ already employed in \cite{yuan}. As already mentioned, for the wind we use the model described in \cite{knigge}. In this case the wind cooling is written as
\begin{equation}
Q^{-}_{\text{win}}=\frac{1}{2}(\eta_b+\eta_k f^2)KR^{\xi+2}\Omega_K^2
\end{equation}
where $f$ is a velocity parameter defined as the ratio of the Keplerian velocity over the escape velocity. Therefore we will set it as $f=\sqrt{2}$. Moreover, $\eta_b$ and $\eta_k$ are efficiency parameters: for $l<\sqrt{3/2}$: $\eta_b=3-2 l^2$ and $\eta_k=1$; and for $l>\sqrt{3/2}$: $\eta_b=0$ and $\eta_k=1-2 f^{-2}(l^2-3/2)$, see \cite{knigge} for details.

Now we have completed the main equations necessary to describe the thermal stability of the flow. Before moving on to close this section and start the stability analysis, let us introduce some useful relations. In what follows we assume that the radius of the disk is $R_d$. In this case the total mass-loss rate, i.e., $\dot{M}_{\text{w}}(R_d)$ can be obtained from equation \eqref{mw} as
\begin{equation}
\dot{M}_{\text{w}}(R_d)=\frac{4\pi K}{s}(R_d^s-R_*^s)
\end{equation}
where the new parameter $s$ is defined as $s=2+\xi$. One may use this equation to find $K$ and normalize $\dot{m}_{\text{w}}(R)$ as follows
\begin{equation}
\dot{m}_{\text{w}}(R)=\frac{s \dot{M}_{\text{w}}(R_d)}{4\pi}\frac{R^{s-2}}{R_d^2-R_*^s}
\end{equation}
As another useful equation, we rewrite the continuity equation \eqref{mah1} as follows
\begin{equation}
\dot{M}_{\text{acc}}(R)=\dot{M}_{\text{acc}}(R_*)+\frac{\dot{M}_{\text{w}}(R_d)}{R_d^s-R_*^s}(R^s-R_*^s)
\label{mah2}
\end{equation}
As the last note, it should be mentioned that the magnetic field is present in our model. Therefore in order to construct a complete set of equations to describe the unknown functions, we need one more constraint on the magnetic field. To do so, we follow the description introduced in \cite{yuan}. Based on MHD simulations \cite{Machida}, one may assume that the strength of the magnetic field decreases by vertical height from the disk midplane. Therefore we simply assume that
\begin{equation}
B_{\phi} H^{\gamma}=\text{constant}=\Phi_{\gamma}
\label{b1}
\end{equation} 
where $\gamma$ is a constant parameter. This equation combined with the contribution of the magnetic field in the pressure content of the flow characterizes the impact of the magnetic field on the evolution of the thermal instability.

\section{Thermal Instability}\label{ins}
  \begin{figure*}
\centerline{\includegraphics[width=5.5cm]{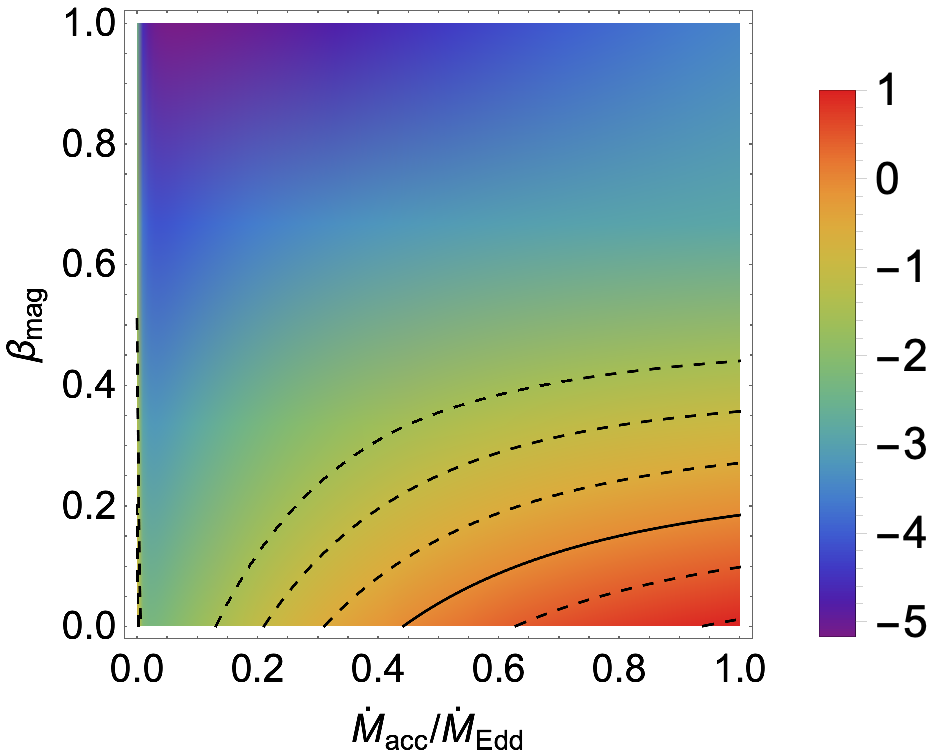}\hspace{0.01 cm} \includegraphics[width=5.5cm]{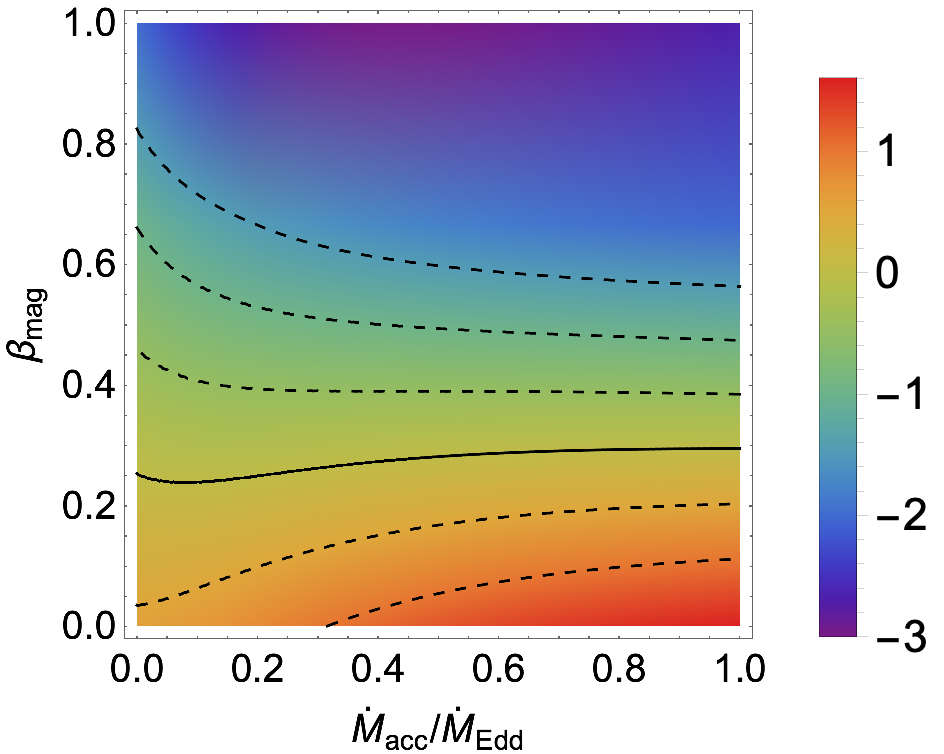}\hspace{0.01 cm} \includegraphics[width=5.5cm]{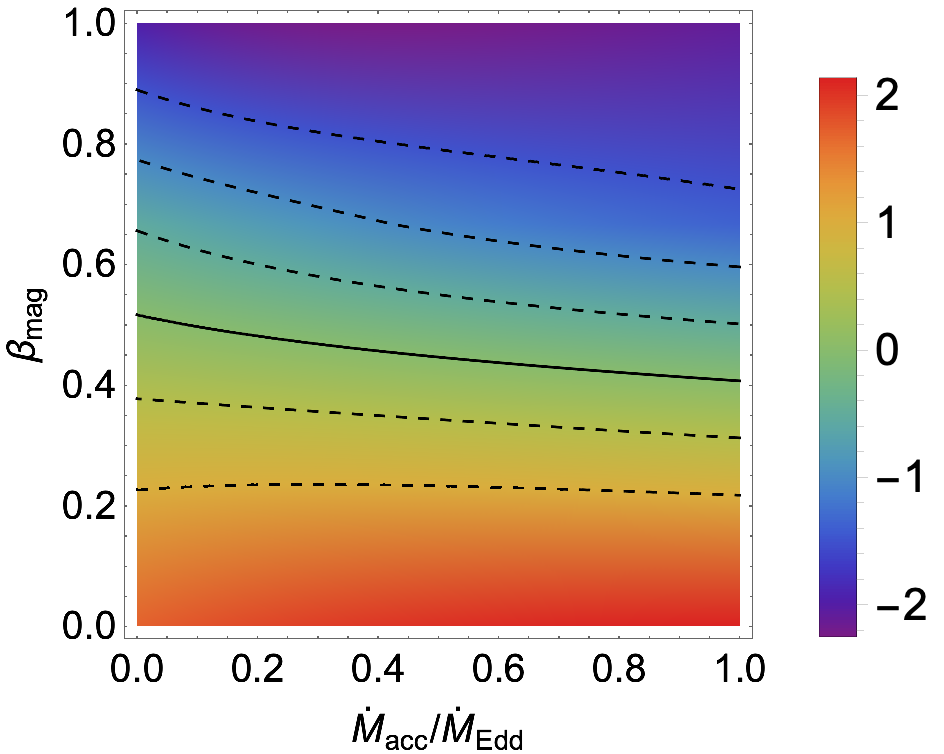}}\vspace{0.5 cm} 
\centerline{\includegraphics[width=5.5cm]{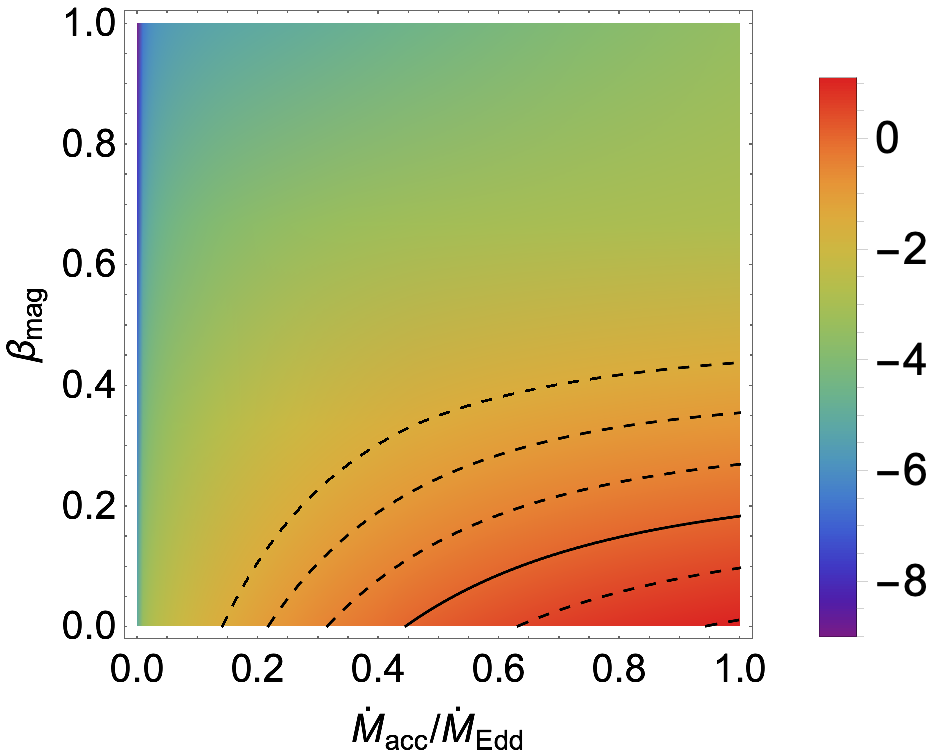}\hspace{0.01 cm} \includegraphics[width=5.5cm]{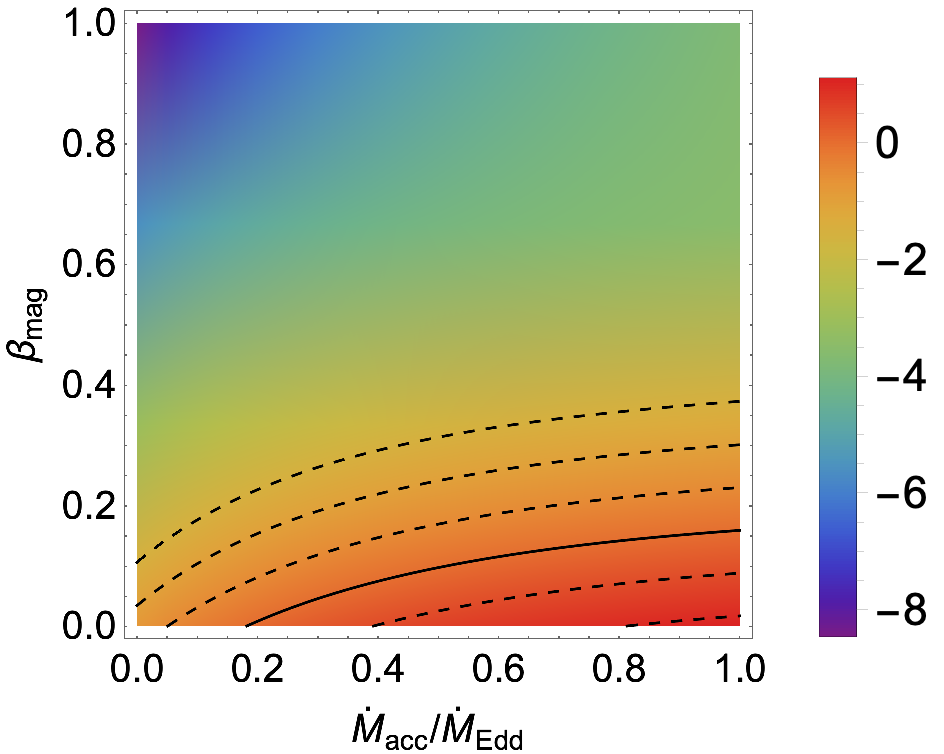}\hspace{0.01 cm} \includegraphics[width=5.5cm]{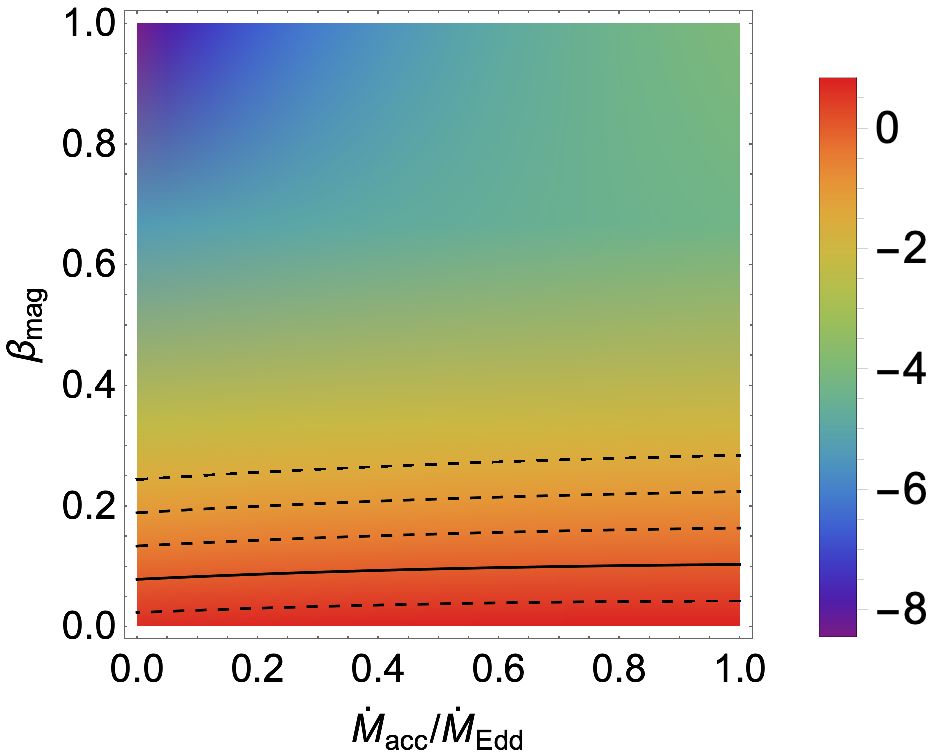}}
\caption{{Top three panels belong to $l^2=2$ and in the bottom panels $l^2=3$. In both rows from left to right we the dimensionless wind accretion rate is chosen as $\dot{M}_{\text{w}}/\dot{M}_{\text{Edd}}=0.001$, $0.1$, $0.5$ respectively. In all panels the curves indicates contours of $\psi=1$, $0.5$, $0$, $-0.5$, $-1$ and $-1.5$ from bottom to up. The solid curve shows the stability boundary $\psi=0$.  Furthermore $s=0.3$ in all panels.}}
\label{fig1}
\end{figure*}
In this section we use the main equations explained in the previous section and find a general criterion for the thermal stability of the disk in the presence of wind. We generalize the analysis of \cite{yuan} to include wind cooling. To do so let us assume that the thermal instability does not change the surface density at the given radius $R$. Note that we are interested to find the stability criterion at an arbitrary radius $R$. Consequently, using equation \eqref{ptot} we find
\begin{equation}
\begin{split}
d \ln p_{\text{tot}}&\simeq d \ln H=\beta_{\text{gas}}(d\ln T-d \ln H)\\&+4(1-\beta_{\text{gas}}-\beta_{\text{mag}})d\ln T+2\beta_{\text{mag}} d\ln B_{\phi}
\end{split}
\label{t1}
\end{equation}
where $\beta_{\text{mag}}=p_{\text{mag}}/p_{\text{tot}}$, $\beta_{\text{gas}}=p_{\text{gas}}/p_{\text{tot}}$ and $\beta_{\text{rad}}=p_{\text{rad}}/p_{\text{tot}}$ are dimensionless quantities. It is clear that 
\begin{equation}
\beta_{\text{mag}}+\beta_{\text{gas}}+\beta_{\text{rad}}=1
\label{t2}
\end{equation}
The wind impact does not appear directly in equation \eqref{t1}. On the other hand, using the energy equation \eqref{q1} we find
\begin{equation}
\begin{split}
d\ln Q_{\text{vis}}^{+} -& d\ln (Q_{\text{rad}}^{-} + Q_{\text{adv}}^{-} +Q_{\text{win}}^{-})=\\& -4 (1- f_{\text{adv}}- f_{\text{win}}) d\ln T+ d\ln T_{{R \Phi}}\\&- f_{\text{adv}}(d\ln {\dot{M}_{\text{acc}}}+ 2 d\ln H)
\end{split}
\label{t3}
\end{equation}
where $f_{\text{adv}}$ and $f_{\text{win}}$ are defined as
\begin{equation}
f_{\text{adv}}=\frac{Q_{\text{adv}}^{-}}{Q_{\text{vis}}^{+} }, ~~~~~~f_{\text{win}}=\frac{Q_{\text{win}}^{-}}{Q_{\text{vis}}^{+} }
\end{equation}
it is clear that wind effects directly appear in equation \eqref{t3}. Moreover, using equations \eqref{av4} and \eqref{b1} we find 
\begin{equation}
d \ln \dot{M}_{\text{acc}}=d \ln T_{R\phi}= d \ln p_{\text{tot}}+d \ln H
\label{t4}
\end{equation}
\begin{equation}
d \ln B_{\phi}=-\gamma d \ln H
\label{t5}
\end{equation}
using equations \eqref{t1} and \eqref{t5} we eliminate $B_{\phi}$ to obtain
\begin{equation}
d \ln p_{\text{tot}}= d \ln H=\frac{4-3\beta_{\text{gas}}-4\beta_{\text{mag}}}{1+2\gamma \beta_{\text{mag}}+\beta_{\text{gas}}}d\ln T
\label{t6}
\end{equation}
Now in order to find the thermal stability criterion in the presence of wind, we substitute equations \eqref{q1}, \eqref{t4} and \eqref{t6} into equation \eqref{t3}. The result is
\begin{equation}
\begin{split}
&\Big[\frac{\partial (Q^{+}_{\text{vis}}-Q^{-}_{\text{rad}}-Q^{-}_{\text{adv}}-Q^{-}_{\text{win}})}{\partial T}\Big]_{\Sigma}\frac{T}{Q_{\text{vis}}^{+} }=\\&~~~~~~~~~~~~~~\frac{\psi}{1+2\gamma\beta_{\text{mag}}+\beta_{\text{gas}}}
\end{split}\label{t7}
\end{equation}
where $\psi$ is defined as
\begin{equation}
\begin{split}
\psi=&4-10 \beta_{\text{gas}}-8(1-\gamma)\beta_{\text{mag}}-12 f _{\text{adv}}+16 f_{\text{adv}}\beta_{\text{gas}}\\&+(16+8\gamma)\beta_{\text{mag}}f_{\text{adv}}+4 f_{\text{win}}(1+2\gamma \beta_{\text{mag}}+\beta_{\text{gas}})
\end{split}\label{t8}
\end{equation}
As we know the thermal instability occurs when 
\begin{equation}
\Big[\frac{\partial (Q^{+}_{\text{vis}}-Q^{-}_{\text{rad}}-Q^{-}_{\text{adv}}-Q^{-}_{\text{win}})}{\partial T}\Big]_{\Sigma}>0
\end{equation}
On the other hand the denominator of the right hand side of \eqref{t7} is positive. Therefore for thermal instability, the nominator should be positive, namely $\psi>0$. In order to study this criterion in a quantitative way for the given parameters $s$, $M$, $R$, $l$, $\gamma$, $\dot{M}_{\text{acc}}(R_*)$ and $\dot{M}_{\text{w}}(R_d)$, we solve the governing equations \eqref{totp}, \eqref{av4}, \eqref{q1}, \eqref{mah2} and \eqref{b1} for six unknowns $\Sigma$, $H$, $T$, $B_{\phi}$, $\dot{M}_{\text{acc}}(R)$ and $\Phi_{\gamma}$. It is clear that we need one more algebraic equation to construct a complete set of equations. To do so we use $\beta_{\text{mag}}=\text{const}$. Finally it is straightforward to plot $\psi$ for the given physical quantities. In fact, we followed the method introduced in \cite{yuan}.

\section{Results}\label{Result}
Now let us investigate the thermal stability of the system concerning changes in the relevant physical quantities. There are several parameters which can influence the stability. For example, it turns out that the accretion rates, magnetic pressure, and wind parameters $l$ and $s$ play a significant role in the local stability of the disk.

Interestingly, the response of the system is sensitive to the magnitude of $l$. More specifically, $l^2=5/2$ is a threshold and the systems behaves completely different for $l^2>5/2$ and $l^2<5/2$. There is a simple interpretation of the existence of this threshold. We already mentioned that when $l^2>3/2$ then $\eta_b=0$ and $\eta_k=5/2-l^2$ (note that $f=\sqrt{2}$). On the other hand, $\eta_k$ directly controls the sign of wind cooling $Q_{\text{win}}^{-}$. When $l^2<5/2$ we have $Q_{\text{win}}^{-}>0$. This means that this type of wind heats the disk. Consequently, in this case, we expect destabilizing behavior for the wind. In contrary, when $l^2>5/2$ the wind cooling is positive and stabilizes the disk.

\subsection{Stability function $\psi(\dot{M}_{\text{acc}},\beta_{\text{mag}})$ for different $\dot{M}_{\text{w}}$}
 In Fig. \ref{fig1}, we have plotted $\psi$ as a function of $\dot{M}_{\text{acc}}$ and $\beta_{\text{mag}}$. In the top and bottom panels, we have set $l^2=2$ and $l^2=3$ respectively. In all the figures we have $\gamma=1$ and $R=10 MG/c^2$. Furthermore, contours indicate curves with constant $\psi$. The solid curve is the stability boundary $\psi=0$. In both rows $\dot{M}_{\text{w}}$ increases from left to right. It is clear that in the top row, by increasing the wind accretion rate, the stability region gets smaller. In other words, when $l^2<5/2$, the existence of wind destabilizes the disk in the sense that higher values for $\beta_{\text{mag}}$ is required to stabilize the disk. One should note that this value for $l$ corresponds to a specific form of wind.

On the other hand, we see in the bottom row that by increasing the wind accretion rate the stability zone gets wider. Strictly speaking, in this case we need lower $\beta_{\text{mag}}$ to stabilize all the accretion rates. Albeit one should note for intermediate values for the wind accretion rate, disks with low accretion rates which are already stable, get thermally unstable. It is important to mention that for both rows by increasing $\beta_{\text{mag}}$ the disk gets stabilized. This result is in agreement with those obtained in \cite{yuan}. In other words, we also confirm that regardless of the nature of the wind, the magnetic field induces stabilizing effects. Of course one should note that we have considered the magnetic field and the wind to be totally independent. This is a restriction for the parametric model adopted here. More specifically, even the power-low mass-loss rate used here may not explain the real winds. This means that a full dynamical model would be needed to find more reliable results. In this regard, this parametric model should be considered as an approximative method which reveals some important features for the system.

Another important feature is that, except the right panels in both rows, when $\beta_{\text{mag}}$ is small the line $\beta_{\text{mag}}=const$ intersects the stability boundary curve $\psi=0$ in two points. This means that in this interval for mass accretion rate the disk is thermally unstable, while for values of mass accretion rate outside this interval the disk is stable. This behavior for the effect of the total accretion rate is also reported in \cite{yuan}. However, we would mention that for one of the points we have $\dot{M}_{\text{acc}}> \dot{M}_{\text{Edd}}$, which belong to slim disk solutions. On the other hand, thin disk solutions are of interest in this study. Therefore we have truncated most of the figures at $\dot{M}_{\text{acc}}=\dot{M}_{\text{Edd}}$.

\subsection{Stability function $\psi(\dot{M}_{\text{acc}},\dot{M}_{\text{w}})$ for different $\beta_{\text{mag}}$}
\begin{figure*}
\centerline{\includegraphics[width=5.5cm]{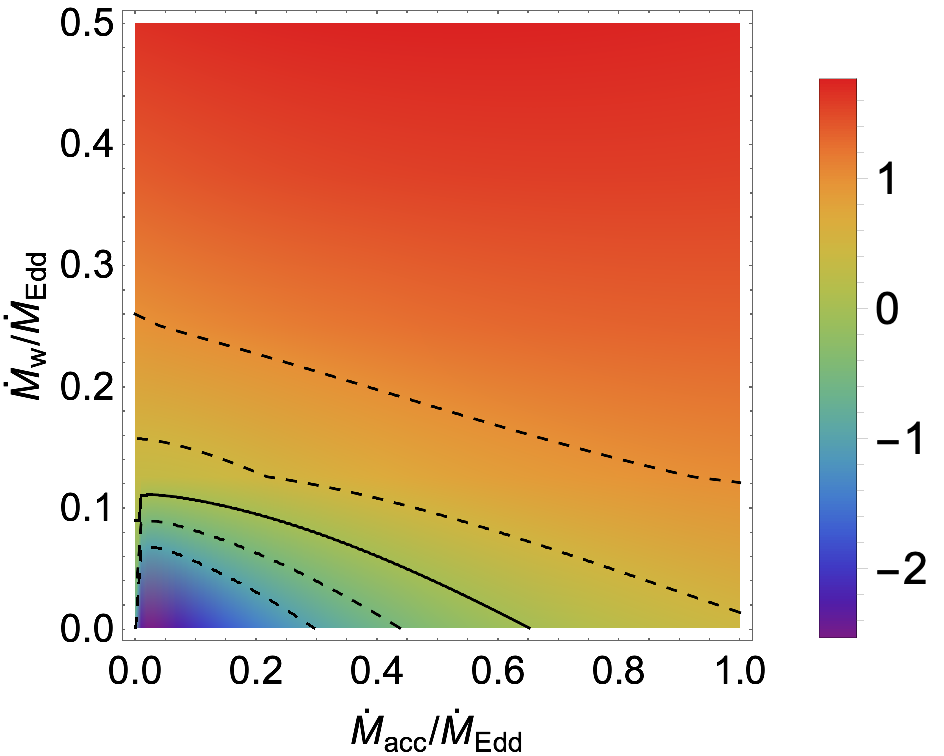}\hspace{0.01 cm} \includegraphics[width=5.5cm]{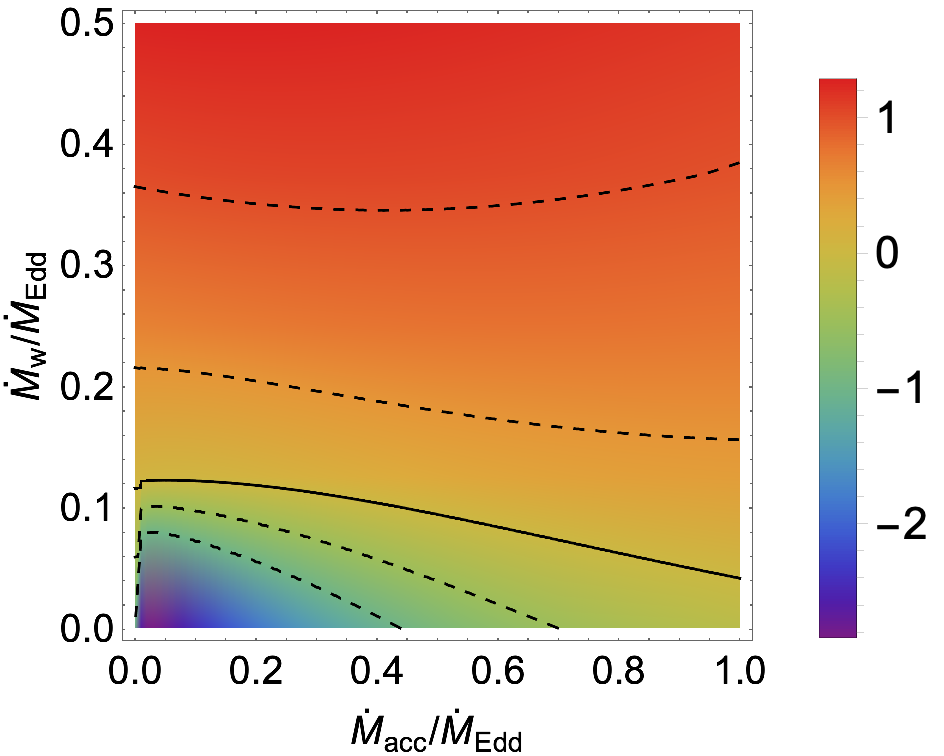}\hspace{0.01 cm} \includegraphics[width=5.5cm]{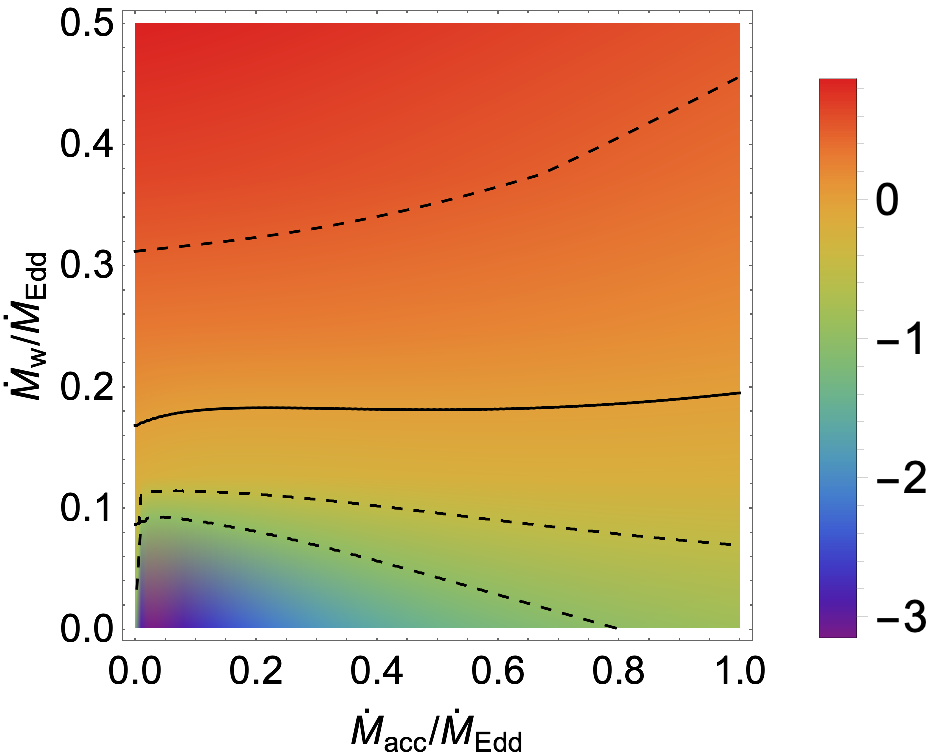}}\vspace{0.5 cm} 
\centerline{\includegraphics[width=5.5cm]{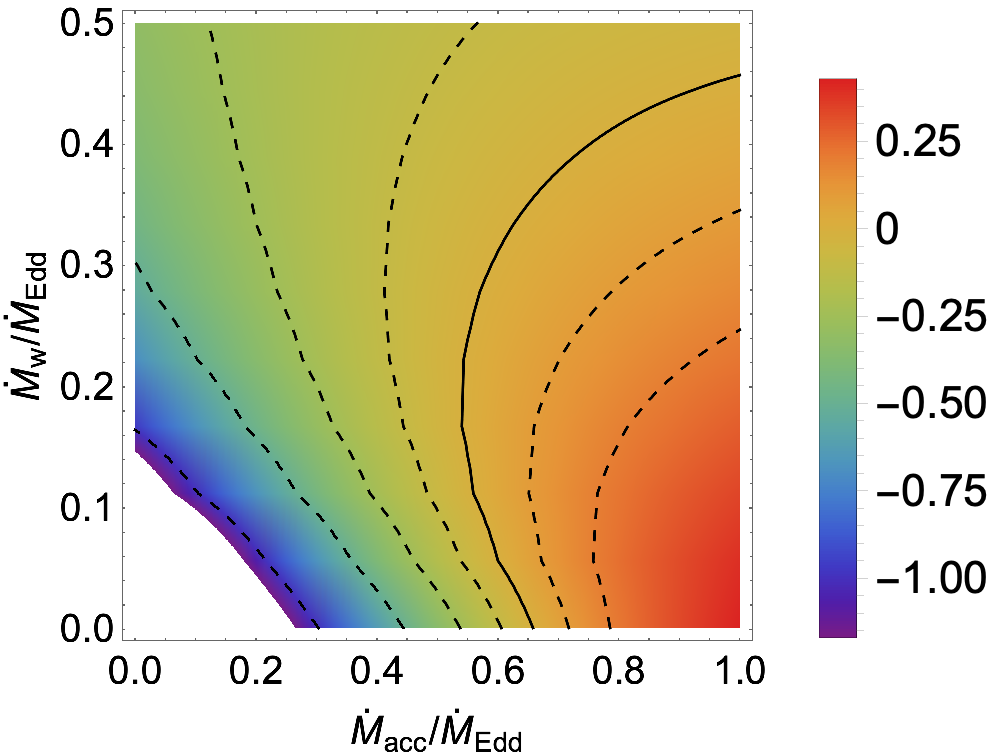}\hspace{0.01 cm} \includegraphics[width=5.5cm]{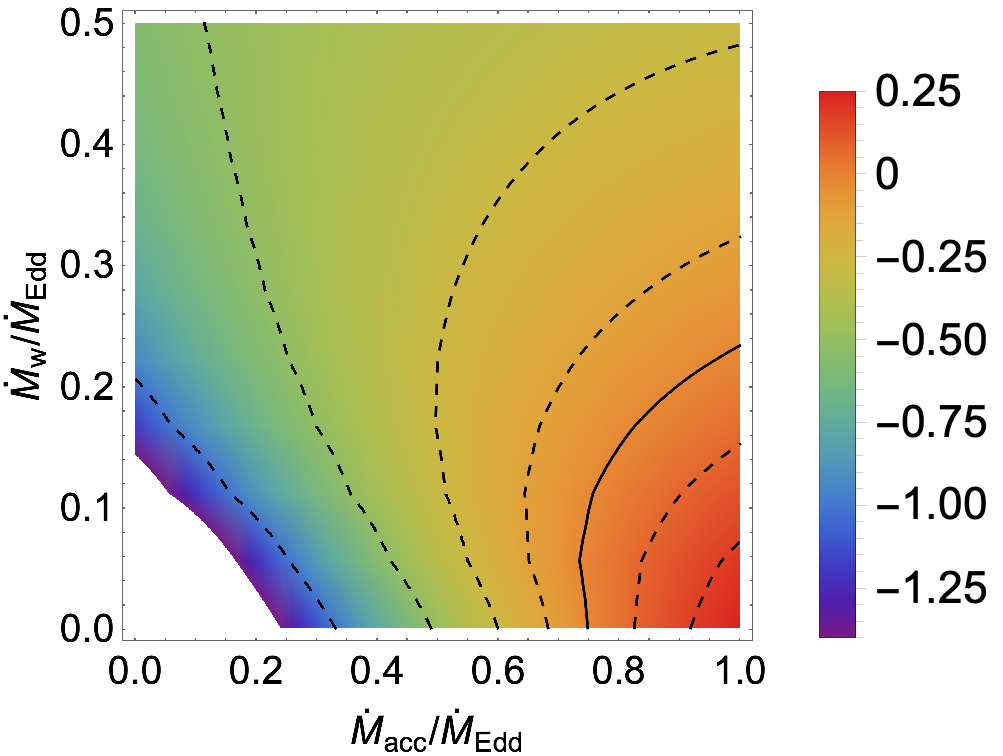}\hspace{0.01 cm} \includegraphics[width=5.5cm]{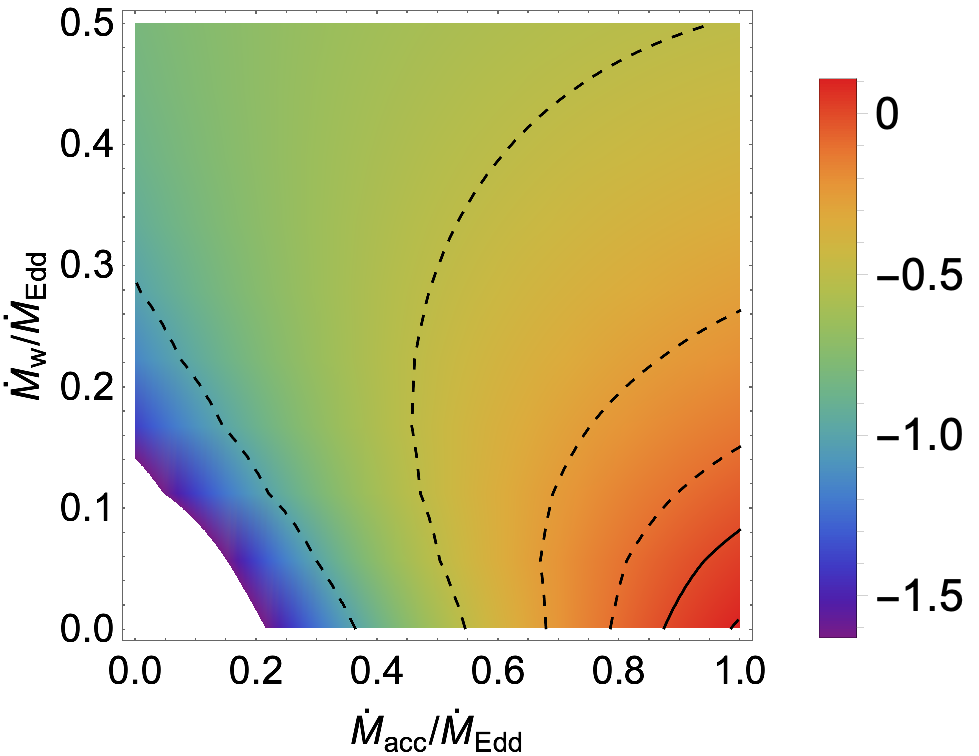}}
\caption{{Top three panels belong to $l^2=2$ and in the bottom panels $l^2=3$. In both rows from left to right we $\beta_{\text{mag}}$ is chosen as $0.1$, $0.125$ and $0.15$ respectively. In the top row the contours indicate $\psi= 1$, $0.5$, $0$, $-0.5$ and $-1$ curves, while for the bottom panel we have $\psi= 0.2$, $0.1$, $0$, $-0.1$ , $-0.25$, $-0.5$ and $-1$. The solid curve shows the stability boundary $\psi=0$.  Furthermore $s=0.3$ in all panels.}}
\label{fig2}
\end{figure*}

In a similar way, it is helpful to plot the stability function $\psi$ as a function of $\dot{M}_{\text{acc}}$ and $\dot{M}_{\text{w}}$, see Fig. \ref{fig2}. In both rows,  from left to right we vary $\beta_{\text{mag}}$ as $\beta_{\text{mag}}=0.1$, $0.125$ and $0.15$. As before, the top row belongs to $l^2=2$ and in the bottom panel we have $l^2=3$. Furthermore the solid curves indicate the stability boundary, i.e., $\psi=0$. 

Let us start with the top panel. It is clear that by increasing the magnetic pressure contribution the stability region gets wider. Also if we keep the mass accretion rate constant and increase the wind accretion rate, we see that the disk eventually gets unstable. Furthermore, if we keep the wind accretion rate constant and move in along the mass accretion rate axis,  at least for relatively small $\dot{M}_{\text{w}}$, we see that the line $\dot{M}_{\text{w}}=const$ intersects the stability boundary in two different $\dot{M}_{\text{acc}}$ (as mentioned above, we discard high mass accretion rates).This directly means that the disk is stable for small and large mass accretion rates. While it is unstable for intermediate values of $\dot{M}_{\text{acc}}$.

Now let us discuss the bottom row in Fig. \ref{fig2}. We recall that in this figure $l^2=3$. Completely in agreement for our discussion for Fig. \ref{fig1}, we see that if we keep the mass accretion rate constant and move along the wind accretion rate, the disk gets stable. As already mentioned, for $l^2>5/2$,  we expect that the existence of the wind stabilizes the disk. On the other hand, as expected magnetic pressure has stabilizing effects and its overall role does not depend on $l$.In other words, we see that the instability region, which covers large mass accretion rates, gets smaller and smaller. This means that the. magnetic pressure stabilizes the. disks with high mass accretion rates and low $\dot{M}_{\text{w}}$. As we see, there is a region in low $\dot{M}_{\text{w}}$ and $\dot{M}_{\text{acc}}$ where our parametric model cannot produce physical accreting disk solutions. This region is not affected by the magnetic pressure. 
\subsection{Stability function $\psi(s, \dot{M}_{\text{w}})$ for different mass accretion rates}

\begin{figure}
\begin{center}
\includegraphics[width=7.5cm]{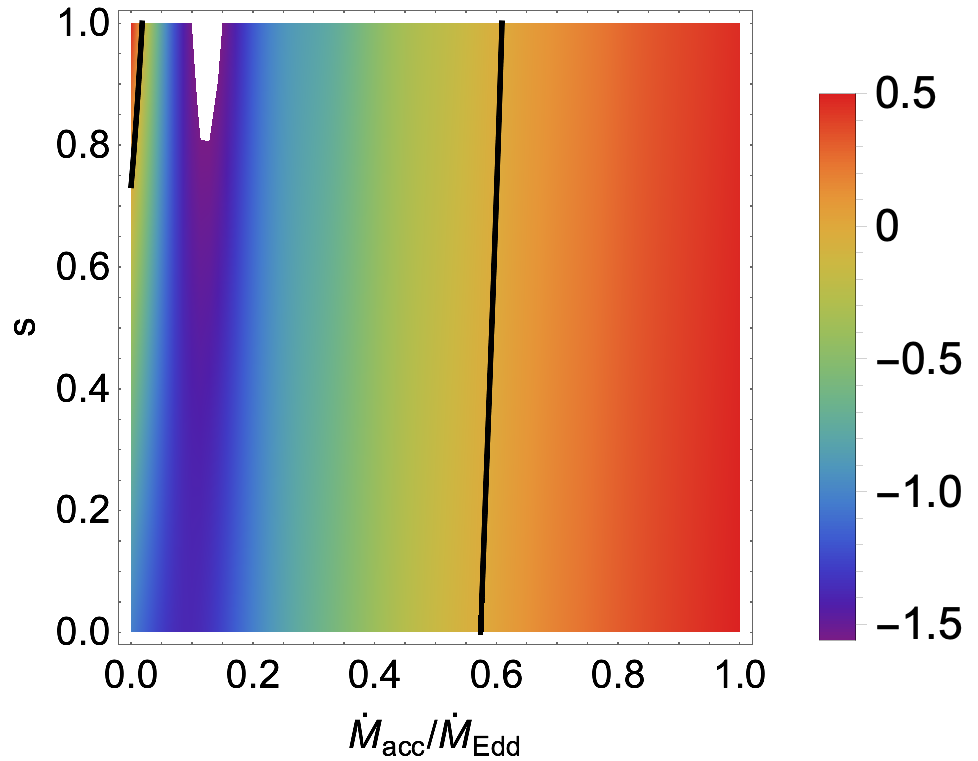}\vspace{0.01 cm} \includegraphics[width=7.5cm]{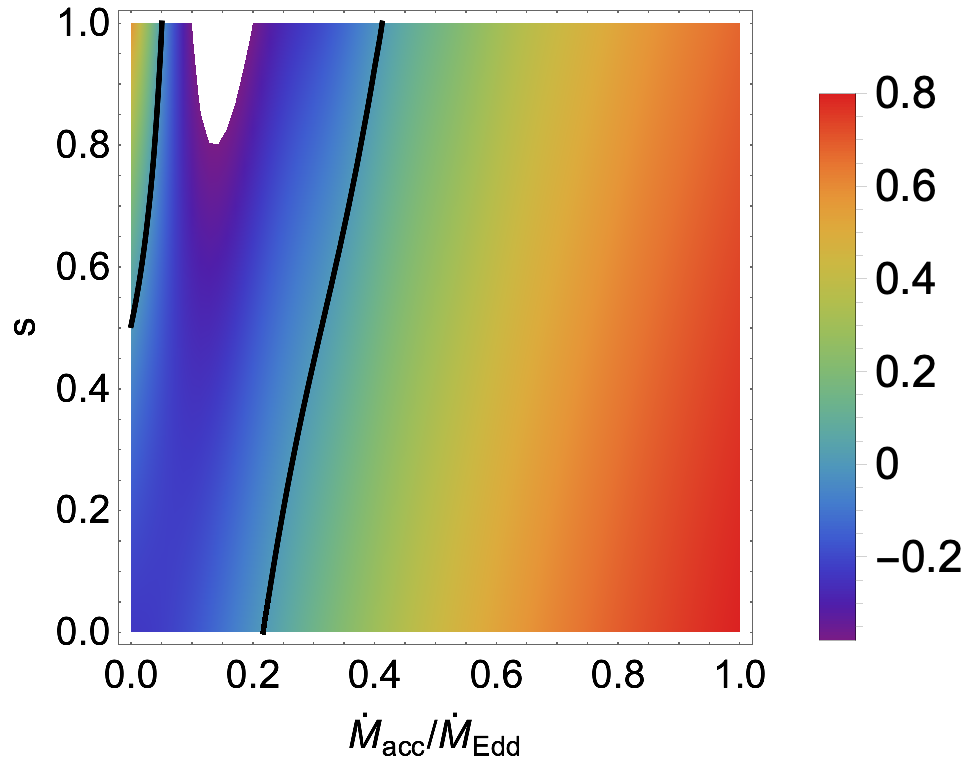}\vspace{0.01 cm} \includegraphics[width=7.5cm]{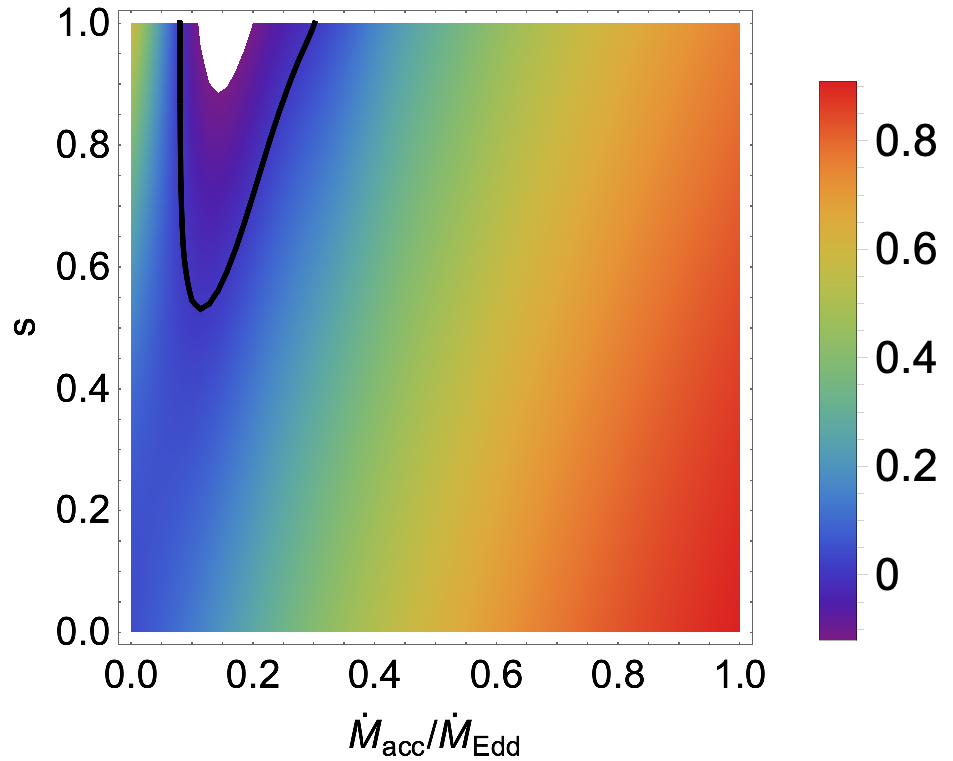}
\end{center}
\caption{{The stability function $\psi$ with respect to $s$ and $\dot{M}_{\text{acc}}$ when $l^2=2$. From up to down the dimensionless wind accretion rate is chosen as $\dot{M}_{\text{w}}/\dot{M}_{\text{Edd}}=0.01$, $0.05$, $0.07$ respectively. In all panels the solid curves indicates the stability contour $\psi=0$.}}
\label{fig3}
\end{figure}

$s$ is another important parameter directly related to the properties of the wind in the system. Therefore it is useful to investigate the response of the system to changes in this parameter. To do so, we have plotted the stability function $\psi(s, \dot{M}_{\text{w}})$ as a density plot in Fig. \ref{fig3}.

From up to down we increase the wind mass accretion rate as $\dot{M}_{\text{w}}/\dot{M}_{\text{Edd}}=0.01$, $0.05$ and $0.07$. In this case, for larger values of the wind accretion rates the disk is completely unstable and the $s$ parameter is not helpful. On the other hand, it is important to mention that in this case, the behavior of the system in the $s-\dot{M}_{\text{w}}$ is not significantly sensitive to the magnitude of $l$. Therefore we have illustrated only the $l^2=2$ case. Furthermore the magnetic pressure contribution is fixed as $\beta_{\text{mag}}=0.1$. The solid black curves separates the stability and instability regions. As already mentioned for this case, by increasing the wind mass accretion rate, the stability region gets smaller. It is clear from the top panel that when wind accretion rate is small, $s$ parameter does not have any impact on the local stability of the system. We see that the mass accretion rate plays a key role. Albeit there is a narrow region for low mass accretion rates which gets unstable when $s\rightarrow 1$.

We see that there is a region around $0.1\dot{M}_{\text{Edd}}<\dot{M}_{\text{acc}}<0.2\dot{M}_{\text{Edd}}$ which does not lead to physical solutions with $s>0.8$. From the middle and bottom panels one may deduce that the parameter $s$ has stabilizing effects. More specifically, although $s$ cannot stabilize the disks with large mass accretion rates $\dot{M}_{\text{acc}}$, it can stabilize low mass accretion rates.

\subsection{Local thermal equilibria: $\dot{M}_{\text{acc}}-\Sigma$ diagram}

\begin{figure*}
 \begin{center}
\includegraphics[width=7cm]{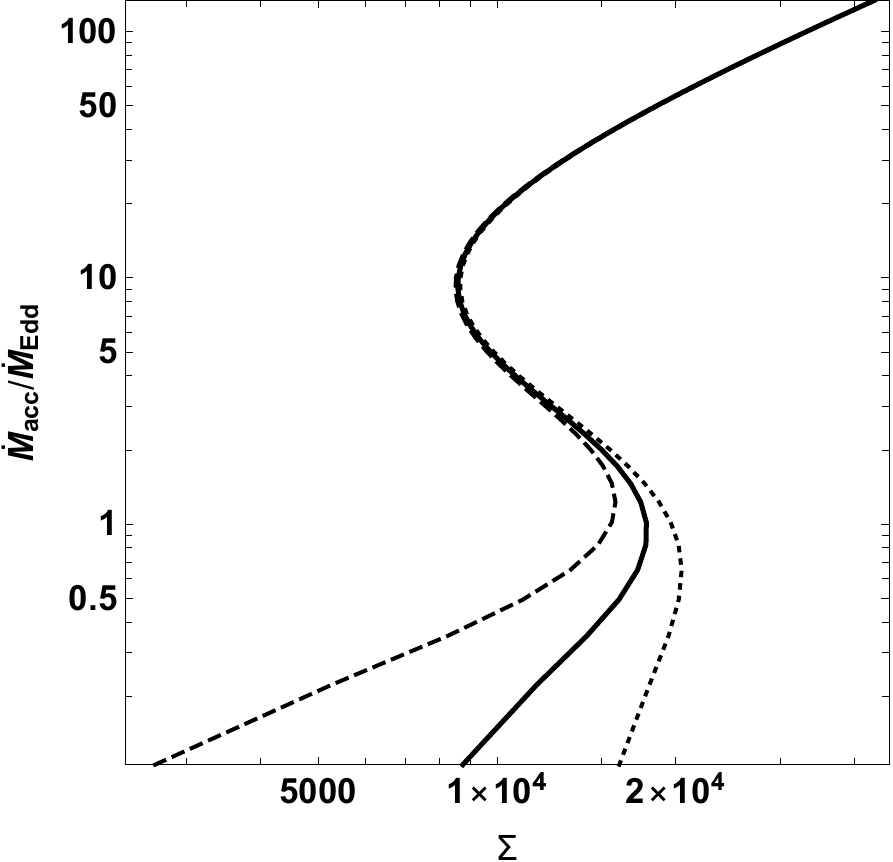}\hspace{0.5 cm} \includegraphics[width=7cm]{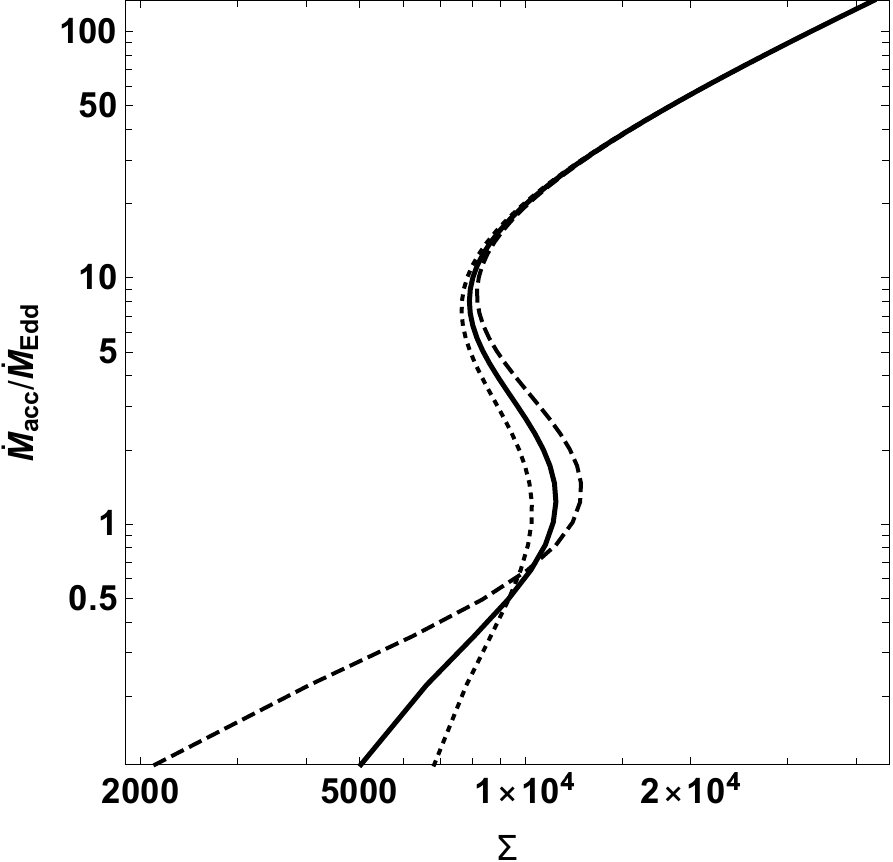}\vspace{0.5cm}
\includegraphics[width=7cm]{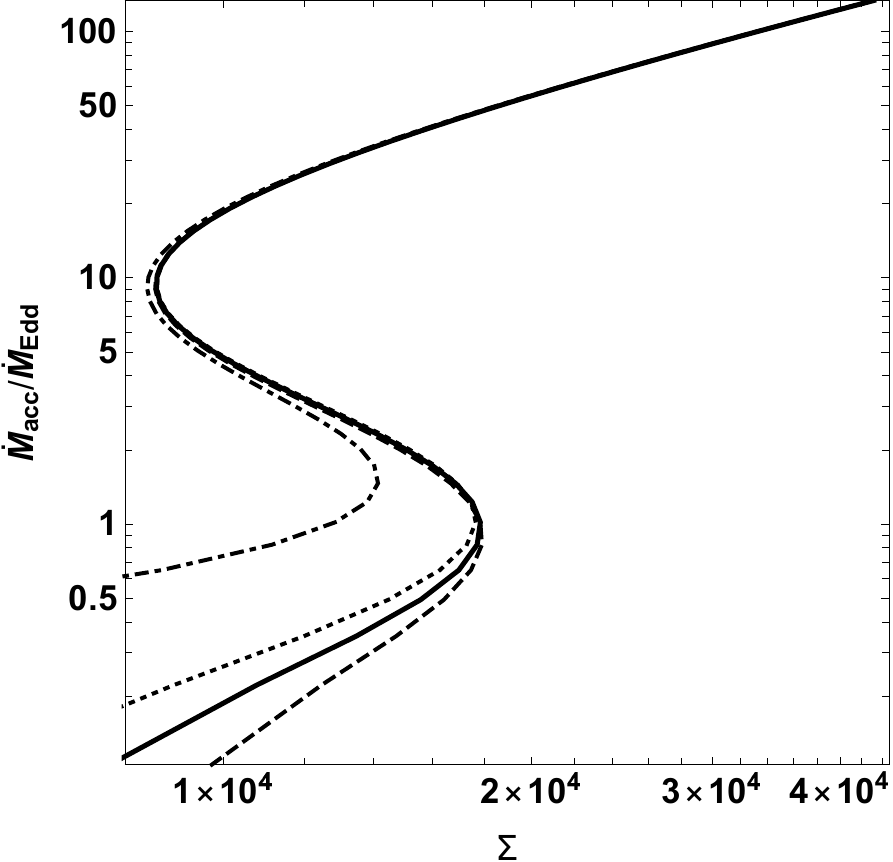}\hspace{0.5 cm} \includegraphics[width=7cm]{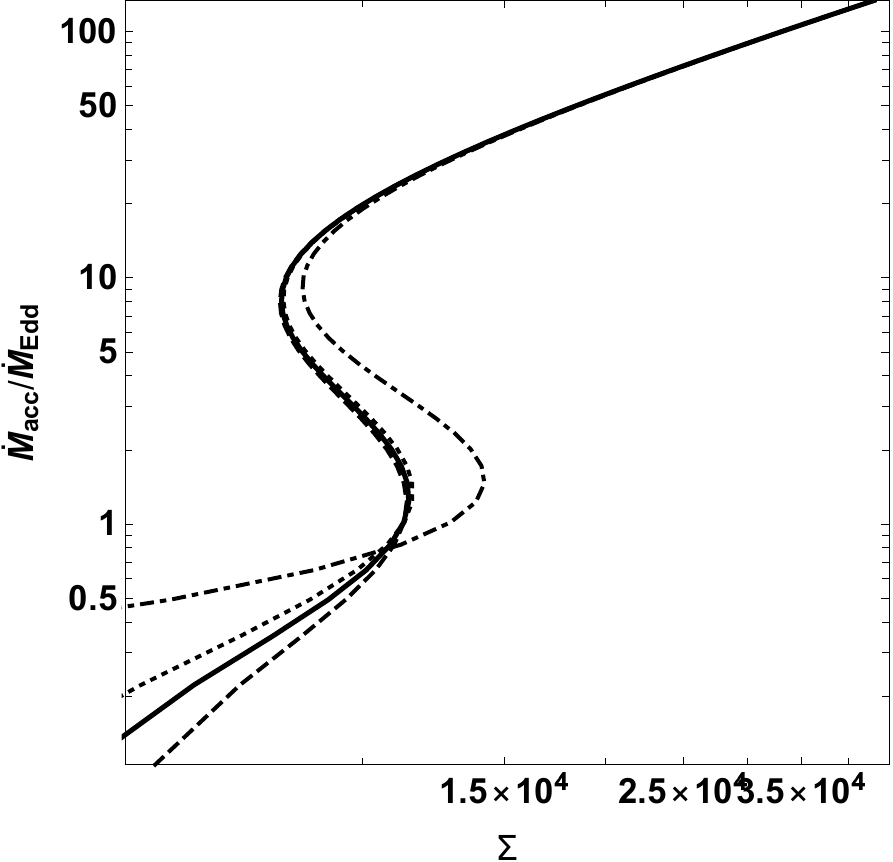}
\end{center}
\caption{{In the left panels we have $l^2=2$ and in the right panels $l^2=3$. In the top panels $s$ is fixed as $s=0.3$ and $\gamma=1$ and $\Phi=3\times 10^{13}$. In the top panels the dashed, solid and dotted curves indicate $\dot{M}_{\text{w}}/\dot{M}_{\text{Edd}}=0.05$, $0.1$ and $0.15$. On the other hand in the bottom panels the dashed, thick, dotted and dot-dashed curves indicate $s=0.2$, $0.4$, $0.6$ and $10$. In this panel the wind accretion rate is constant $\dot{M}_{\text{w}}=0.1$.}}
\label{fig4}
\end{figure*}

It is also convenient and instructive to plot the $\dot{M}_{\text{acc}}-\Sigma$ diagram for the local stability of the system. In this diagram negative slop shows the thermally unstable solution. We have plotted this diagram in Fig. \ref{fig4}. It is crucial to mention again that only accretion rates smaller than the Eddington critical rate, i.e. $\dot{M}_{\text{acc}}\lesssim \dot{M}_{\text{Edd}}$, is of interest to us. However, only for the sake of better presentation, we have included high accretion rates in Fig. \ref{fig4}. In this case we will be able to see the well-known $S$-shape stability curve.

 Let us first discuss the top row. In the left panel, we have $l^2=2$ and in the right panel $l^2=3$. To see the response of the system to changes in $l$ and wind accretion rate, in both panels we keep $s$ constant ($s=0.3$). Also other relevant parameters are $\gamma=1$ and $\Phi=3\times 10^{13}$. In this row, the dashed, solid and dotted curves indicate $\dot{M}_{\text{w}}/\dot{M}_{\text{Edd}}=0.05$, $0.1$ and $0.15$. It is clear from the top left panel when $l^2<5/2$ by increasing the contribution of the wind, the disk gets unstable for a wider range of $\dot{M}_{\text{acc}}$ and $\Sigma$. On the other hand, in the top right panel when $l^2>5/2$, we see that by increasing $\dot{M}_{\text{w}}$ the slop of $\dot{M}_{\text{acc}}-\Sigma$ curves gets negative in a smaller interval. This directly means that in this case, the existence of wind induces stabilizing effects. These results are completely in harmony with what we already saw in our density plots of $\psi$. 

 Similarly, let us discuss the bottom row in Fig. \ref{fig4}. In the left and right panels, we have $l^2=2$ and $l^2=3$ respectively. In this row, we are interested to check the response of the disk to $s$. To do so, we keep the wind accretion rate as $\dot{M}_{\text{w}}=0.1$ and vary the parameter $s$. We recall that other parameters are chosen as $\gamma=1$ and $\Phi=3\times 10^{13}$. The dashed, thick, dotted and dot-dashed curves indicate $s=0.2$, $0.4$, $0.6$ and $10$. We see that when $s$ is small the stability of the system is not significantly sensitive to $s$ in both cases, i.e., $l^2<5/2$ and $l^2>5/2$. On the other hand, when $s$ is relatively large, we see in the bottom left panel that the parameter $s$ causes stabilizing effects. For example $s=10$ stabilizes all the equilibrium solutions with $\dot{M}_{\text{acc}}\lesssim 2\dot{M}_{\text{Edd}}$. Accordingly, one may conclude from the bottom right panel that $s$ has destabilizing effects. One should note that by increasing this parameter, the slope of the diagram gets negative for a wider interval of $\Sigma$. This behavior was not clear in our previous figures dealing with the stability function $\psi$. We recall that $s$ is not completely independent of $\dot{M}_{\text{w}}$. However, its behavior is opposite to wind accretion rate. 

There is another interesting feature in all the panels of Fig. \ref{fig4}. It is clear that the wind cannot influence the thermal stability of disks with high mass accretion rates. We see that the upper stable branch, i.e., $\dot{M}_{\text{acc}}\gtrsim10 \dot{M}_{\text{Edd}}$ in the panels does not change by varying the wind parameters. In this branch of solutions, advective cooling plays a key role and the radiation pressure gets very large. Both of these quantities have stabilizing nature dominating the dynamics of the system. Therefore it is natural that in this case both magnetic pressure and wind do not have a serious impact on the system. On the other hand, as we already discussed, wind effects significantly influence the lower stable branch and the unstable interval.

As a final remark in this subsection, we have also explored the behavior of the special case $l=1$. As we already mentioned, in this case, $C_{w}=0$. We found that thin disk with $l=1$ does not show any different behavior compared to $1<l^2<5/2$. In other words, this type of wind with $l=1$ destabilizes the disk.
\section{Summary and discussion}\label{sum}
In this paper, we have studied the thermal stability of thin accretion disks. We present a full local stability analysis when the disk is magnetized and there is a wind mechanism in the system. We recall that the thermal stability of radiation dominated thin disks has revealed some puzzles. For example, in the standard picture, the radiation dominated thin disk is thermally unstable when the mass accretion rate is higher than a few percents of Eddington rate.  However, observations of the X-ray binaries do not confirm this prediction. We have revised this issue by including the role of the wind in the system. We found a criterion for the stability of the disk when bot the magnetic field and wind exist in the disk, see equation \eqref{t8} ($\psi>0$). For the magnetic field, we use the key assumption already used in \cite{yuan}, namely the magnetic field will become weaker whit increase of the scale height $H$ or the temperature $T$. However, our focus is on the wind's effects and we used a wind model already introduced in \cite{knigge}. We showed that depending on the type of the wind, the disk can be stabilized or destabilized. In other words, when the wind parameter $l^2<5/2$ then the existence of the wind makes the disk unstable. On the other hand, for $l^2>5/2$, the wind significantly stabilizes the disk. Interestingly this stabilizing effect is completely in agreement with those presented in \cite{2014ApJ...786....6L}. In fact, it is shown in \cite{2014ApJ...786....6L} that the critical accretion rate, corresponding to the thermal instability threshold, is significantly increased in the presence of the magnetic driven disk winds. On the other hand, as we already mentioned, when $l>1$ in the parametric wind model presented in \cite{knigge}, then the model is suitable for describing the magnetic driven disk winds. This means that our analysis, also confirms the stabilizing role of the magnetic driven wind reported in \cite{2014ApJ...786....6L}. In some senses, this consistency shows that Knigge's parametric wind model is a viable one at least for the magnetic driven disk case. However, we should mention that the parametric approach presented here, is an approximative procedure. Strictly speaking, the real wind mechanisms may not follow a simple model presented by \cite{knigge}. As discussed in \cite{knigge}, a full dynamical model is required to obtain an accurate description. However, this simple approximative model is useful to investigate some main features of the thin accretion disks. In fact, we have shown in this paper,  that Knigge's parametric model, is useful to investigate the thermal stability of the disks. Another restriction is that we have ignored the direct effects of the outflow on the magnetic filed, e.g. via field line stretching. A full analysis removing the above-mentioned restrictions, is beyond the scope of this paper and we leave it foe future studies.

It is important mentioning that when there is no wind, to stabilize the disk, a large contribution for the magnetic pressure compared to the total pressure is required ($\beta_{\text{mag}}>0.2$). While this contribution is larger than the typical value $\beta_{\text{mag}}\simeq 1$ widely used in relevant simulations. Therefore, to stabilize the disk with normal values of magnetic pressure, one needs to assume $\gamma>3$, see \cite{yuan} for more details. This value gets even larger when the stability of the radiation dominated thin disk simulated in \cite{2009ApJ...704..781H} has been considered. To explain the stability of this thin disk with magnetic field one needs $\gamma>7$. These large values for $\gamma$ seem uncomfortably as reported in \cite{yuan}. 

However, when wind exists in the disk, we can achieve stability for any value of the mass accretion rate. In this case it is not required to assume a large value for $\beta_{\text{mag}}$ or $\gamma$. Of course, in this case, one needs to justify the existence of wind in thin disks.

\acknowledgments
We would like to thank Mahmood Roshan for helpful comments and discussions.

\bibliographystyle{apj}

\end{document}